\documentclass[conference,final]{IEEEtran}

\usepackage{amsmath}
\usepackage[utf8]{inputenc}
\usepackage{amsfonts}
\usepackage{amssymb}
\usepackage{amsthm}
\usepackage{amsbsy}
\usepackage{pgfplots}
\usepackage{epstopdf}
\usepackage{tikz}
\usepackage{pgfplots}
\usepackage{textpos}
\usepackage{balance}
\usetikzlibrary{dsp,chains}
\usetikzlibrary{shapes,arrows}
\usepackage{mathtools}
\usepackage{color}
\usepackage{graphicx}
\usepackage{bm}
\usepackage{soul}

\usepackage{epstopdf}

\usepackage[acronym,shortcuts]{glossaries}

\newacronym{CSI}{CSI}{channel state information}
\newacronym{iid}{iid}{independent and identically distributed}
\newacronym{MIMO}{MIMO}{multiple-input-multiple-output}
\newacronym{MIMOME}{MIMOME}{multiple-input-multiple-output-multiple-eavesdropper}
\newacronym{ML}{ML}{maximum likelihood}
\newacronym{MSE}{MSE}{mean square error}
\newacronym{NA}{NA}{no attack}
\newacronym{PCA}{PCA}{pilot contamination attack}
\newacronym{PSK}{PSK}{phase shift keying}
\newacronym{SK}{SK}{secret key}
\newacronym{SNR}{SNR}{signal to noise ratio}
\newacronym{SVD}{SVD}{singular value decomposition}
\newacronym{SOP}{SOP}{secrecy outage probability}

\tikzstyle{decision} = [diamond, draw, 
    text width=4.5em, text badly centered, node distance=3cm, inner sep=0pt]
\tikzstyle{block} = [rectangle, draw, 
    text width=15em, text centered, rounded corners, minimum height=4em]
\tikzstyle{line} = [draw, -latex']

\newcounter{MYtempeqncnt}
\newcounter{MYtempeqncnt2}

\IEEEoverridecommandlockouts

\begin{document}

\title{Pilot Contamination Attack Detection by Key-Confirmation  in Secure MIMO Systems}
\author{Stefano Tomasin\thanks{S. Tomasin is also with University of Padova, Italy.}, Ingmar Land and Fr\'ed\'eric Gabry \\ \small Mathematical and Algorithmic Sciences Lab, France Research Center,\\ Huawei Technologies Co. Ltd.\\
emails: firstname.lastname@huawei.com}
\maketitle

\begin{abstract}
Many security techniques working at the physical layer need a correct \ac{CSI} at the transmitter, especially when devices are equipped with multiple antennas. Therefore such techniques are vulnerable to \acp{PCA} by which an attacker aims at inducing false \ac{CSI}. In this paper we provide a solution to some \ac{PCA} methods, by letting two legitimate parties to compare their channel estimates. The comparison is made in order to minimize the information leakage on the channel to a possible attacker. By reasonable assumptions on both the channel knowledge by the attacker and the correlation properties of the attacker and legitimate channels we show the validity of our solution. An accurate analysis of possible attacks and countermeasures is provided, together with a numerical evaluation of the attainable secrecy outage probability when our solution is used in conjunction with beamforming for secret communications.
\end{abstract}

\glsresetall

\section{Introduction}
In  order to enhance security in communication systems, it has been recently proposed to operate at the physical layer (see \cite{7270404} and references therein for an overview). In particular, since early Shannon works \cite{Shannon}, is has been shown that, by properly coding a message it is possible to perfectly recover it at an intended receiver without leaking any information on it to an eavesdropping device. The use of devices with multiple antennas in what is called \ac{MIMOME} system, in general increases the security capabilities. However full knowledge of the legitimate channel (\ac{CSI}) is needed at the transmitter to exploit at best these potentials \cite{1504.07154v1}. For example, if  Alice beamforms the signal to the legitimate receiver Bob, the eavesdropper Eve will (most likely) see a poor channel, and therefore the resulting gap between the \acp{SNR} at Bob and Eve will be large\footnote{Note that if the channel to Eve is known, perfect secrecy can be achieved. When CSI to the eavesdropper is not available, its statistics may still be available and then we can assess the probability that Eve obtains information on the transmitted message.}. 

In order to obtain \ac{CSI} when channel reciprocity is available, Bob can send pilot signals in a training phase and Alice will estimate the channel. However, a \ac{PCA} has been outlined in the literature \cite{1504.07154v1, 6151778}, working as follows. Pilot symbols are assumed publicly known; therefore when Bob transmits pilots, Eve may transmit the same pilots as well. The channel estimated by Alice will then be the sum of the channels between Bob and Eve. Therefore, with this attack, Eve is able to modify the channel estimate and obtain a higher \ac{SNR} after beamforming. 
 
A number of solutions have been proposed in the literature to detect \ac{PCA}:
in \cite{6666096} it is proposed to reserve two slots for training; in the second slot random pilots on a \ac{PSK} constellation are transmitted and the receiver can detect the pilot contamination by correlating the received signals in the two phases and checking if the resulting correlated signal belongs to a \ac{PSK} constellation. 
The use of an additional action after channel estimation, in which Alice beamforms a pilot sequence to Bob such that the received signal at Bob has predetermined average/statistical values is advocated in \cite{7136233}. If the estimated channel is not correct, the beamformer will not correspond to the channel and the parameters of the signal received by Bob will not have the values agreed upon. In \cite{7248525} a solution to the pilot contamination problem has been proposed, under the hypothesis that the legitimate users exactly know the correlation matrix of both the legitimate and the eavesdropper channel, which may be unrealistic in many application scenarios. In \cite{6675552} it is proposed to detect unexpected drops in the received signal strength due to the misaligned beam direction. However, this approach requires a reference \ac{SNR} level in the absence of the attack, which is difficult to obtain, especially when multiple legitimate users are present. A method in \cite{6831343} allows Alice to detect \ac{PCA}, based on the fact that she will observe a change in the statistics of training, i.e., an increase of the power of training, due to the transmission by Eve. However, also in this case a good \ac{SNR} reference value is needed. 

In this paper we propose a technique by which Alice and Bob estimate the channel by exchanging pilot sequences and then check if their estimates coincide. In order to avoid disclosing the channel to Eve and to prevent Eve from forging the message exchange, the comparison is not done publicly. Instead, they extract \acp{SK} from the estimates and then check if they coincide using a key confirmation procedure. This protocol ensures that Eve gets the minimum information on the channel from the information exchange. We consider various scenarios for the channel knowledge by Eve and for the correlation between legitimate's and attacker's channels, assessing the validity of the proposed solution.

The rest of the paper is organized as follows. In Section II we introduce the transmission scenario and we indicate the baseline protocol for channel estimation. In Section III the proposed solution is described. An attack analysis under different scenarios is proposed in Section IV. Numerical results are reported in Section V, before conclusions are drawn in Section VI.

{\em Notation.} Matrices and vectors are indicated in boldface. $\bm{X}^H$ denotes the Hermitian operator of matrix $\bm{X}$, $\bm{I}_N$ is the $N \times N$ identity matrix. $\bm{X}^{-1}$ denotes the inverse of matrix $\bm{X}$. $\log$ denotes base-2 logarithm. ${\det}(\bm{X})$ is determinant of matrix $\bm{X}$.

\section{System Model}

All users are transceivers equipped with $N$ antennas (the proposed methods can be straightforwardly extended to the case in which users have different numbers of antennas) and channels are assumed narrowband, time-invariant and reciprocal. Let $\bm{H}$ be the $N \times N$ \ac{MIMO} channel matrix of complex numbers describing the channel between Alice and Bob, and let $\bm{G}_1$ and $\bm{G}_2$ be the matrices of the Alice-Eve and Bob-Eve channels, respectively. Alice and Bob know the noise power and the statistics of $\bm{H}$.

\begin{figure}
\centering
\begin{tikzpicture}
\node[circle,draw,minimum size=1.5cm] (A) {Alice};
\node[coordinate,right=of A] (c) {};
\node[circle,draw,below= of c,minimum size=1.5cm] (E) {Eve};
\node[circle,draw,right= of c,minimum size=1.5cm] (B) {Bob};
\draw[dspline] (A) -- node[left,above] {$\bm{H}$}  (B); 
\draw[dspline] (A) -- node[right,below,xshift=-3] {$\bm{G}_1$}  (E); 
\draw[dspline] (B) -- node[left,below,xshift=5] {$\bm{G}_2$}  (E); 
\end{tikzpicture}
\caption{Transmission scenario.}
\label{scenario}
\end{figure}
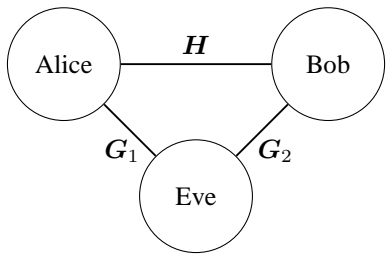

Initially, we assume that no user knows the channel matrices\footnote{Other assumptions on the channel knowledge by Eve will be considered in Section IV.}. Alice aims at securely transmitting a message to Bob  \cite{7248525} or at extracting a \ac{SK} with him \cite{6675552, 6831343}. To this end, Alice estimates her channel to Bob $\bm{H}^{\rm (A)}$ and then performs beamforming with  matrix $\bm{B}$ to Bob's antennas. Assuming that Alice has no knowledge of her channel to Eve, the beamforming procedure aims at reducing the \ac{SOP}, i.e., the probability that Eve knows something about the secret message or key. However, it is pivotal that Alice has a correct \ac{CSI} of the link to Bob, otherwise the beamforming is not effective. 

\noindent \paragraph*{Baseline Channel Estimation Protocol} The baseline channel estimation protocols provides that Bob sends pilot signals to Alice, who estimates the channel $\bm{H}^{\rm (A)}$ and performs beamforming. In her transmission, Alice will also insert pilots, so that Bob can estimate the channel including beamforming and is able to decode the message.  

\subsection{Attacker Capabilities}

The purpose of the attacker Eve is to obtain information on the secret message or key, and in this respect she performs a \ac{PCA} in order to increase the \ac{SOP}. 

We assume that Eve has unlimited transmit power and computation capabilities, she is able to detect and intercept any transmission between Alice and Bob, and she is able to transmit signals in a synchronous manner with them. Moreover, signaling and pilot signals used by Alice and Bob are already known to Eve. Eve will exploit these capabilities to corrupt the \ac{CSI} obtained by Alice to perform beamforming, so that the outage probability is increased. Various levels of knowledge of the channels and their statistics can be assumed, and a detailed analysis will be provided in Section IV.

\noindent \paragraph*{Baseline Attack} A \ac{PCA} strategy provides that Eve transmits pilot signals together with Bob, so that the estimated channel at Alice will be 
\begin{equation}
\label{baselineattack}
\bm{H}^{\rm (A)} = \left[\bm{H} + \bm{G}_1\right] + \bm{w}^{\rm (A)}\,,
\end{equation}
where $\bm{w}^{\rm (A)}$ is the noise matrix of residual estimation error.

\section{Proposed Solution}

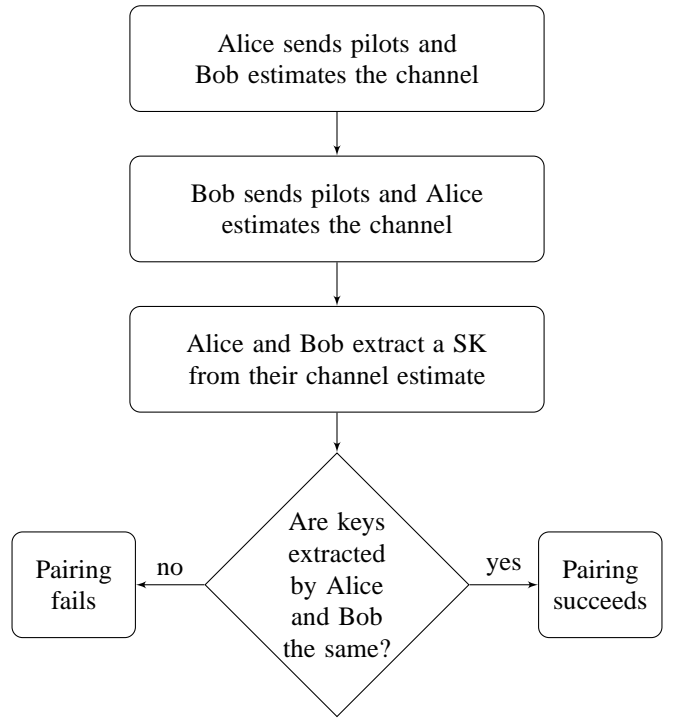
\begin{figure}
\centering  
\begin{tikzpicture}[node distance = 2cm, auto]
    \node [block] (A) {Alice sends pilots and Bob estimates the channel};
    \node [block, below of=A] (B) {Bob sends pilots and Alice estimates the channel };
    \node [block, below of=B] (C) {Alice and Bob extract a \ac{SK} from their channel estimate};
    \node [decision, below of=C] (D) {Are keys extracted by Alice and Bob the same?};
    \node [rectangle, draw, text centered, rounded corners, minimum height=4em, right of=D,text width=4em,xshift=1.5cm] (E) {Pairing succeeds};
    \node [rectangle, draw, text centered, rounded corners, minimum height=4em, left of=D,text width=4em,xshift=-1.5cm] (F) {Pairing fails};

    \path [line] (A) -- (B);
    \path [line] (B) -- (C);
    \path [line] (C) -- (D);
    \path [line] (D) -- node {yes} (E);
    \path [line] (D) -- node [above] {no} (F);
\end{tikzpicture}
\caption{Block diagram of the proposed solution.}
\label{bd_idea}
\end{figure}

In order to prevent the \ac{PCA}, we propose the protocol shown in Fig. \ref{bd_idea}. It comprises two major steps: a) channel estimation, b) channel comparison. 

For the first step, two pilot transmission phases are provided: first Alice transmits pilots to Bob, and then Bob transmits pilots to Alice; the pilots are assumed to be publicly known. Based on the received pilots, the two legitimate users estimate the channel. In the second step, the two channel estimates are compared: if they coincide (apart from the effect of the noise), then Eve did not perform the {\em baseline attack}, while if they differ, Eve was transmitting pilots during the second phase. In fact, when Eve attacks with the baseline attack, she transmits pilots in the second phase, and the two channel estimates will differ. Note that if Eve transmits pilots also in the first phase, the estimated channels at Alice and Bob will be 
\begin{equation}
\bm{H}^{\rm (A)} = \left[\bm{H} + \bm{G}_1\right] + \bm{w}^{\rm (A)}\,,
\end{equation}
\begin{equation}
\bm{H}^{\rm (B)} = \left[\bm{H} + \bm{G}_2\right] + \bm{w}^{\rm (B)}
\end{equation}
and since $\bm{G}_1 \neq \bm{G}_2$, the two channels will differ. 

For a channel comparison, a first trivial solution to compare the two estimates would be that Alice and Bob send the estimates publicly on the wireless channel. However, this would allow Eve to intercept the transmission and then know the channel. Once Eve knows the true channel between Alice and Bob, she can perform a \ac{PCA} in which she precodes the pilots such that both Alice and Bob estimate channel $\bm{G}_1$, thus getting full access to forthcoming transmission. This attack is detailed in Section \ref{attackfull}. 

\paragraph*{Improved Channel Comparison} In order to check the consistency of the two channels without revealing them to Eve, Alice and Bob independently extract a \ac{SK} from the estimated channels (see \cite{Bloch} for details on the \ac{SK} extraction procedure). By this technique they obtain a bit sequence (key) from the channel estimates. Denote $a$ the key generated by Alice and $b$ the key generated by Bob. Let $M$ be the length of the extracted keys (in number of bits).

If no attack is present, the two bit sequences are identical despite noise that has affected the estimation (see \cite{Bloch, 7063620} for details). In the presence of an attack that has modified the channel in a way such that the change of the channel estimates is above the noise level, the two bit sequences (keys) at Alice and Bob will be different.  Therefore a key confirmation procedure is used to check if the two extracted keys coincide. 

\paragraph*{On The Number of Antennas} Note that as the number of antennas at the transceivers increases ($N \rightarrow \infty)$, the channel estimation phase will require more resources (time and energy). However a trade-off between performance and resource utilization can be obtained when antennas are unbalanced among the transceivers. For example, if Alice has a huge number of antennas and Bob only a few, channel Estimation by Alice is simpler than by Bob. In this case Bob sends pilots to Alice, who in turns either selects a subset of antennas or beamforms signals through the antennas and sends pilots to let Bob estimate a reduced equivalent channel.

\subsection{Key Confirmation Procedure}

\begin{figure}
\centering
\begin{tikzpicture}[node distance = 2cm, auto]
    \node [block] (A) {Alice sends a random sequence of bits encrypted with the \ac{SK}};
    \node [block, below of=A] (B) {Bob decodes the random sequence and processes it with an invertible non-identity publicly known function };
    \node [block, below of=B] (C) {Bob sends the new sequence (mapped value) to Alice, by encrypting it with his \ac{SK} };    
    \node [block, below of=C] (D) {Alice decodes the transmitted sequence and decrypts it with her \ac{SK} };
        \node [decision, below of=D] (E) {equal to function of original sequence?};
    \node [rectangle, draw, text centered, rounded corners, minimum height=4em, right of=E,text width=4em,xshift=1.5cm] (F) {Pairing succeeds};
    \node [rectangle, draw, text centered, rounded corners, minimum height=4em, left of=E,text width=4em,xshift=-1.5cm] (G) {Pairing fails};

    \path [line] (A) -- (B);
    \path [line] (B) -- (C);
    \path [line] (C) -- (D);
    \path [line] (D) -- (E);        
    \path [line] (E) -- node {yes} (F);
    \path [line] (E) -- node [above] {no} (G);
\end{tikzpicture}
\caption{Key confirmation procedure.}
\label{keyconf}
\end{figure}
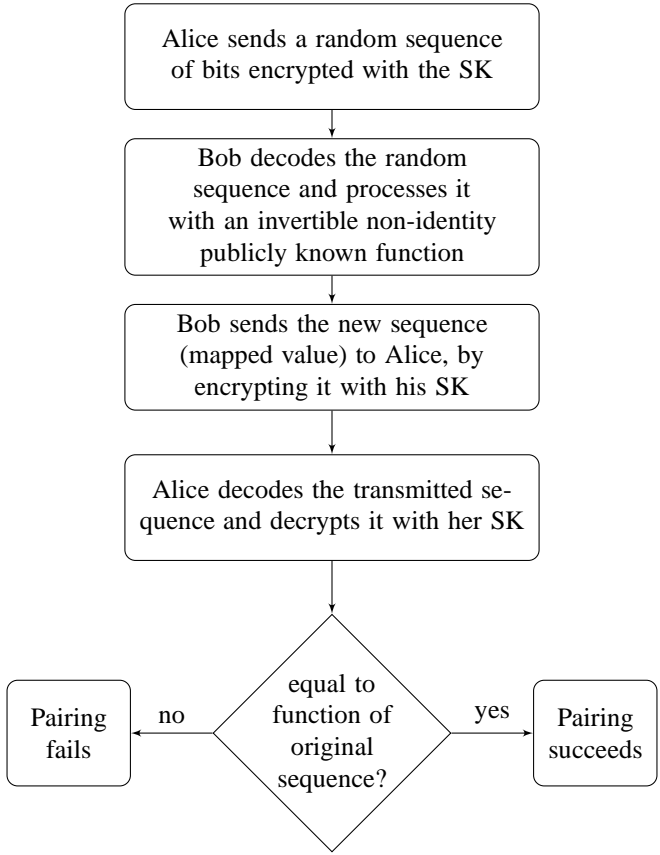

A vast literature on key confirmation procedure is available, see for example \cite{Menezes}. In the following we present a possible approach, summarized in Fig. \ref{keyconf}. Alice generates a sequence of $M$ random bits, denoted by $r$, and xors them with the \ac{SK} (bit-wise modulo-two sum) to obtain $x = a + r$, i.e., Alice performs a one-time-pad encoding of the random bits. The encoded bits $x$ are transmitted over the channel. Bob detects these encoded bits and decrypts them using his extracted key $b$, in order to obtain the random bit sequence generated by Alice: $r’ = x + b$ (modulo-two sum). Note that $r = r'$ if $a = b$. Bob then maps the decrypted bits into another sequence of $M$ bits by applying an invertible non-identical function\footnote{$F_2$ is the binary Galois field.} $h(\cdot): F_2^M \rightarrow F_2^M $ to obtain the mapped value $h(r’)$. On this mapped value Bob performs one time padding with his key and sends the encrypted message to Alice: $y = h(r’) + b$. Alice decrypts the message by removing the one-time pad, $z = y + a$,  and checks if the received message is the correct mapped value of her randomly generated bits, i.e. checks if 
\begin{equation}
h(r) =z.
\label{check}
\end{equation}
If the mapped value and $z$ are identical (i.e., (\ref{check}) holds true), Alice concludes that there is no \ac{PCA}, and pairing is successful. 
A few facts are worth noting: 

$a)$ The condition for the check (\ref{check}) is given by $h(r) = h(x+a)+b$; therefore function $h(\cdot)$ must be non-identical. 

$b)$  When Alice and Bob estimate the same channel, the extracted keys are the same and the whole procedure leads to a successful pairing. If instead the keys extracted by Alice and Bob do not correspond, the decryption by Bob will lead to another message, thus to another mapped value, and the pairing process will fail. 

$c)$ Eve does not learn the channel $\bm{H}$ during this process. 

$d)$ If Eve performs an attack in the phases following the training phase, she does not get to know the \ac{SK} and thus can not break the pairing process.

\subsection{Channel Reciprocity Issues}

In the proposed solution the channel reciprocity between Alice and Bob is very important. However, this is relevant not only for the detection of pilot contamination but also for the forthcoming use of the channel estimate by Alice, i.e., beamforming secret messages to Bob. If the channel is different due to hardware impairments, the channel that will be used by Alice after the pairing process will be not correct. In fact, she will beamform a signal to Bob using the channel estimate, but if this is not correct, the signal will not reach Bob, who will not be able to decode the secret information. So the assumption that hardware impairments are moderate is associated with the use of this estimate. When we look specifically at robustness of our proposed solution to (moderate) mismatches, we note that channel $\bm{H}$ is used to extract the \ac{SK}. Now, the \ac{SK} extraction procedure, available in existing literature includes the fact that Alice and Bob have different estimates of the channel. Therefore, the solution can be considered robust to hardware impairments. Impairments still diminish the protection against attacks, since for a higher noise (or impairments') power, a smaller number of \ac{SK} bits can be extracted from the channel. 

\section{Attack Analysis}

In this section we assess the robustness of the proposed approach against various attack strategies, in particular distinguishing the level of knowledge about the channels that Eve has.

For the analysis we focus on the case in which the number of antennas $N$ tends to infinity, i.e., a massive \ac{MIMO} scenario. We also assume that   the \ac{SK} agreeing procedure is done optimally, i.e., with capacity achieving strategies. Furthermore, we assume that channels $\bm{H}$, $\bm{G}_1$ and $\bm{G}_2$ have \ac{iid} zero-mean Gaussian entries.

We start assuming that Eve does not know a-priori channels $\bm{H}$, $\bm{G}_1$ and $\bm{G}_2$. Let us denote with $\sigma_H^2$ the variance of each entry of $\bm{H}$, and let us assume that channel estimates are affected by {\em estimation noise} ($\bm{w}^{\rm (A)}$ and $\bm{w}^{\rm (B)}$) with power $\gamma$ per entry. Following the results of \cite{6129510}, we have that the \ac{SK} capacity is 
\begin{equation}
C(\rho) = \log \frac{1}{1-\rho^2}\,,
\label{capa}
\end{equation}
where 
\begin{equation}
\rho = \frac{{\rm E}[H^{\rm (A)}_{i,j} H^{\rm (B)*}_{i,j}]}{\sigma_H^2 + \gamma}
\end{equation}
is the normalized correlation coefficient. When \ac{NA} is performed we have 
\begin{equation}
\rho^{({\rm NA})} = \frac{\sigma_H^2}{\sigma_H^2 + \gamma}\,.
\label{rhoNA}
\end{equation}
Now, in the presence of an attack, we can consider the signal transmitted by Eve as additional noise that will decrease the correlation among the observed channels. If we indicate with  $\rho^{({\rm A})}$ the normalized correlation coefficient, we will have $\rho^{({\rm A})} < \rho^{({\rm NA})}$, and from (\ref{capa}) the \ac{SK} capacity of the resulting channel will be reduced. 

Assuming that Alice and Bob implement a \ac{SK} agreement procedure tailored to $C(\rho^{({\rm NA})})$, while the effective \ac{SK} capacity is $C(\rho^{({\rm A})})$, by the strong converse \cite{6129510}, we conclude that the probability that the two distilled keys are different is asymptotically going to 1. Assuming an ideal key confirmation procedure (i.e., error-free and non corrupted by Eve), we can then conclude that the attack is always detected.

\subsection{Attack Strategy with Partial \ac{CSI} at Eve, Uncorrelated Channels and Oblivious Alice}

We now consider the case in which Eve knows her channel to Alice and Bob ($\bm{G}_1$ and $\bm{G}_2$), however, we assume that she does not know $\bm{H}$, thus she has a partial \ac{CSI}. Furthermore, we assume that $\bm{G}_1$ and $\bm{G}_2$ are uncorrelated with $\bm{H}$. Alice and Bob are oblivious of Eve, and in particular they do not know that Eve knows her own channels. In this case Eve can precode pilots in the first phase by precoder $\bm{G}_2^{-1} \bm{Q} $ and in the second phase by precoder $\bm{G}_1^{-1} \bm{Q} $, where $\bm{Q}$ is any $N \times N$ matrix, so that the channel estimated by Alice and Bob is always $\bm{H}^{\rm (A)}= \bm{H}^{\rm (B)}=\bm{H} + \bm{Q}$ (plus noise). Therefore Alice and Bob will distill the same key, and the key confirmation procedure will confirm that this is the channel between the two users. 

Still, note that since Eve does not know the legitimate channel $\bm{H}$, she is not able to induce a specific channel\footnote{For example, Eve cannot impose that the estimated channel by Alice and Bob is the Alice-Eve channel, $\bm{H}^{\rm (A)} = \bm{H} + \bm{Q} = \bm{G}_1$, which would be the most advantageous situation for Eve, who would be able to access to all forthcoming secret transmissions.}. Therefore $\bm{Q}$ will be randomly chosen with \ac{iid} zero-mean entries having variance $\sigma_Q^2$. This choice will yield an observed channel variance $\sigma_H^{(\rm A) 2} = \sigma_Q^2 + \sigma_H^2$, thus providing the correlation when \ac{PCA} is performed by Eve
\begin{equation}
\rho^{({\rm A})} = \frac{\sigma_H^2 + \sigma_Q^2}{\sigma_Q^2 + \sigma_H^2 + \gamma}> \rho^{({\rm NA})}.
\end{equation}

\paragraph*{Attack countermeasure} As a countermeasure to this attack, Alice and Bob can check if the traces of $\bm{H}^{(A)}\bm{H}^{(A)H}$ and $\bm{H}^{(B)}\bm{H}^{(B)H}$ are $N\sigma_H^2$, i.e., these two conditions must be satisfied (for $N \rightarrow \infty$)
\begin{subequations}
\label{tracecon}
\begin{equation}
{\rm trace}(\bm{H}^{(A)}\bm{H}^{(A)H}) = N(\sigma_H^2 +1)
\end{equation}
and
\begin{equation}
{\rm trace}(\bm{H}^{(B)}\bm{H}^{(B)H}) = N(\sigma_H^2 +1) \,.
\end{equation}
\end{subequations}

Clearly, when Eve performs the attack described above, conditions (\ref{tracecon}) will not be satisfied, since by the independence of $\bm{Q}$ with $\bm{H}$ we have 
\begin{equation}
\begin{split}
{\rm trace} &(\bm{H}^{(B)}\bm{H}^{(B)H}) = \\
& {\rm trace}(\bm{H}\bm{H}^{H}) +{\rm trace}(\bm{W}\bm{W}^{H}) + {\rm trace}(\bm{Q}\bm{Q}^H) = \\
& N(\sigma_H^2 +1+\sigma^2_Q) > N(\sigma_H^2 + 1)\,.
\end{split}
\end{equation}

Note that this countermeasure requires that both Alice and Bob know the channel statistics, in particular they must know $\sigma_H^2$. The estimate of this parameter can be done by averaging over a long time the channel estimate,  assuming that other conditions (such as path-loss) do not change and that Eve is not transmitting over this time. Indeed, this countermeasure has the same drawbacks as the approach in \cite{6831343}.

\subsection{Attack Strategy with Partial \ac{CSI},  Correlated Channels and Oblivious Alice}
\label{corrsec}

Here we assume that $\bm{G}_1$ and $\bm{G}_2$ are correlated with $\bm{H}$, and that Eve knows this statistic. Moreover, Eve knows the realization of $\bm{G}_1$ and $\bm{G}_2$, but she does not know $\bm{H}$, thus having a partial \ac{CSI}. In particular, by defining the column vectors $\bm{h}$, $\bm{g}_1$ and $\bm{g}_2$ obtained by vectorization of the columns of $\bm{H}$, $\bm{G}_1$ and $\bm{G}_2$, respectively, the correlation matrices are 
\begin{equation}
\bm{K}_1 = {\rm E}[\bm{g}_1 \bm{h}^H] \quad \bm{K}_2 = {\rm E}[\bm{g}_2 \bm{h}^H]\quad \bm{R}_{12} = {\rm E}[\bm{g}_1 \bm{g}_2^H]
\end{equation}
\begin{equation}
\quad \bm{R}_1 = \bm{R}_2 = {\rm E}[\bm{g}_1 \bm{g}_1^H] = {\rm E}[\bm{g}_2 \bm{g}_2^H] = \sigma_G^2 \bm{I}_{N^2}\,.
\end{equation}

Alice and Bob ignore the knowledge by Eve. Therefore, the \ac{SK} agreement procedure is still be based on (\ref{capa}), using as correlation factor $\rho^{(\rm NA)}$ of (\ref{rhoNA}). As we have seen from the previous attack case, a successful attack must not decrease the correlation coefficient $\rho$ with respect to $\rho^{\rm NA}$, and at the same time it must satisfy trace constraints (\ref{tracecon}). Since Eve knows both $\bm{G}_1$ and $\bm{G}_2$ she can precode pilots by $\bm{G}_2^{-1} \bm{Q} $ and $\bm{G}_1^{-1} \bm{Q}$ in the first and second phase, respectively. In this way, Alice and Bob will estimate the same channel and therefore the correlation among channels will be unaltered.

Now $\bm{Q}$ can be chosen on the basis of Eve's knowledge of the channel, and in order to pass the trace check (\ref{tracecon}). In particular, let $\bm{H}^{\rm (E)}$ be the \ac{ML} estimate of the channel $\bm{H}$ available at Eve, which under Gaussian assumptions corresponds to the minimum \ac{MSE} estimator. Again, let us stack into the column vectors $\bm{h}^{\rm (E)}$ and $\bm{q}$ the matrices $\bm{H}^{\rm (E)}$ and $\bm{Q}$, respectively. Then the vector $\bm{x} = [\bm{h}^T \bm{g}_1^T \bm{g}_2^T]^T$ is zero-mean Gaussian with correlation matrix
\begin{equation}
\bm{R} = {\rm E}[\bm{x}\bm{x}^H] = \left[\begin{matrix}
\sigma_H^2\bm{I}_{N^2} & \bm{K}_1^H & \bm{K}_2^H \\
\bm{K}_1 & \sigma_G^2\bm{I}_{N^2} & \bm{K}_1 \bm{K}_2^H \\
\bm{K}_2 & \bm{K}_2\bm{K}_1^H & \sigma_G^2\bm{I}_{N^2} \\
\end{matrix}\right]\,. 
\end{equation}
By defining
\begin{equation}
\bm{S} = \bm{R}^{-1} = \left[\begin{matrix}
\bm{S}_{11} & \bm{S}_{12} & \bm{S}_{13} \\
\bm{S}_{21} & \bm{S}_{22} & \bm{S}_{23} \\
\bm{S}_{31} & \bm{S}_{32} & \bm{S}_{33} 
\end{matrix}\right] 
\end{equation}
the \ac{ML} estimator at Eve is 
\begin{equation}
\bm{h}^{\rm (E)} = -[\underbrace{\bm{S}_{11}^{-1}\bm{S}_{12}}_{\bm{A}} \bm{g}_1 +\underbrace{\bm{S}_{11}^{-1} \bm{S}_{13}}_{\bm{B}} \bm{g}_2]\,.
\end{equation}

The optimal attack strategy by Eve provides that she tries to let both Alice and Bob estimate channel $\bm{g}_1$ instead of $\bm{h}$, so that Eve gets the most favorable channel conditions. Therefore she will choose
\begin{equation}
\label{expq}
\bm{q} = -\bm{h}^{\rm (E)} + \alpha \bm{g}_1\,,
\end{equation}
where $\alpha$ is a scaling factor chosen to satisfy the trace constraint. In particular, we impose
\begin{equation}
\begin{split}
{\rm trace} &[(\bm{h} + \bm{q})(\bm{h} + \bm{q})^H] \underset{N \rightarrow \infty}{=} {\rm trace}\{{\rm E}[(\bm{h} + \bm{q})(\bm{h} + \bm{q})^H]\} \\
& = N^2\sigma_H^2
\end{split}
\label{tracecheck}
\end{equation}
\setcounter{MYtempeqncnt}{\value{equation}}where we have indicated the asymptotic behavior ($N \rightarrow \infty$) of the trace. Now we have (\ref{lunga}) at the top of next page. By the \ac{MSE} orthogonality principle we have $\bm{M}_2 = \bm{0}$, therefore (\ref{tracecheck}) is satisfied with \addtocounter{equation}{1}
\begin{equation}
\alpha = \frac{\sqrt{N^2\sigma_H^2 - {\rm trace}(\bm{M}_1)}}{N}\,.
\end{equation}

\paragraph*{Remark} We note that the derived $\bm{Q}$ provides also the optimal attack when the \ac{SK} procedure is not performed, but the baseline channel estimation protocol described in Section II is adopted. Indeed, the choice of $\bm{Q}$ as detailed improves over the baseline attack (\ref{baselineattack}).

\begin{figure*}
\setcounter{MYtempeqncnt2}{\value{equation}}
\setcounter{equation}{\value{MYtempeqncnt}}
\begin{equation}
\begin{split}
{\rm E}  [(\bm{h} + \bm{q})(\bm{h} + \bm{q})^H] = &{\rm E}[\bm{h}\bm{h}^H] +{\rm E}[\bm{h}\bm{q}^H] + {\rm E}[\bm{q}\bm{h}^H] + {\rm E}[\bm{q}\bm{q})^H]  = \\
& \sigma_H^2\bm{I}_{N^2} + {\rm E}[\bm{h}((\bm{A} + \alpha\bm{I}_{N^2}) \bm{g}_1)^H + \bm{h}(\bm{B}\bm{g}_2)^H] + {\rm E}[(\bm{A} + \alpha\bm{I}_{N^2}) \bm{g}_1\bm{h}^H + (\bm{B}\bm{g}_2)\bm{h}^H] + \\
& +{\rm E}[((\bm{A} + \alpha\bm{I}_{N^2})\bm{g}_1 + \bm{B}\bm{g}_2 )((\bm{A} + \alpha\bm{I}_{N^2})\bm{g}_1 + \bm{B}\bm{g}_2 )^H] = \\
& = \sigma_H^2 \bm{I}_{N^2} + \bm{K}_1 (\bm{A}^H + \alpha\bm{I}_{N^2}) + \bm{K}_2\bm{B}^H  + (\bm{A} + \alpha\bm{I}_{N^2})\bm{K}_1^H + \bm{B}\bm{K}_2^H +\\
&+ (\bm{A} + \sigma_G^2\alpha\bm{I}_{N^2}) (\bm{A} + \alpha\bm{I}_{N^2})^H + (\bm{A} + \alpha\bm{I}_{N^2})\bm{R}_{12} \bm{B}^H +  \bm{B}\bm{R}_{12}^H(\bm{A} + \alpha\bm{I}_{N^2})^H + \sigma_G^2 \bm{B}\bm{B}^H = \\
& = \sigma_H^2 \bm{I}_{N^2} + \bm{K}_1 \bm{A}^H + \alpha\bm{K}_1  + \bm{K}_2\bm{B}^H +\bm{A} \bm{K}_1^H + \alpha \bm{K}_1^H + \bm{B}\bm{K}_2^H +\\
&+ \sigma_G^2 \bm{A}  \bm{A} + \alpha \bm{A} + \alpha \bm{A}^H + \alpha^2\bm{I}_{N^2} +\bm{A} \bm{R}_{12} \bm{B}^H + \alpha \bm{R}_{12} \bm{B}^H + \alpha  \bm{B}\bm{R}_{12}^H + \bm{B}\bm{R}_{12}^H \bm{A} + \sigma_G^2\bm{B}\bm{B}^H = \\
& = \underbrace{[\sigma_H^2 \bm{I}_{N^2} + \bm{K}_1 \bm{A}^H + \bm{K}_2\bm{B}^H  +\bm{A} \bm{K}_1^H + \bm{B}\bm{K}_2^H +\bm{A} \bm{R}_{12} \bm{B}^H + \sigma_G^2\bm{A} \bm{A} + \bm{B}\bm{R}_{12}^H \bm{A} + \sigma_G^2\bm{B}\bm{B}^H]}_{\bm{M}_1} + \\
& \alpha \underbrace{[\bm{K}_1  +  \bm{K}_1^H  + \bm{A}+ \bm{A}^H +  \bm{R}_{12} \bm{B}^H +   \bm{B}\bm{R}_{12}^H]}_{\bm{M}_2} + \alpha^2\bm{I}_{N^2} 
\end{split}
\label{lunga}
\end{equation}
\setcounter{equation}{\value{MYtempeqncnt2}}
\hrulefill
\vspace*{4pt}
\end{figure*}

\subsection{Attack Strategy with Full Channel Knowledge and Oblivious Alice}
\label{attackfull}

Clearly, if Eve also knows the legitimate channel $\bm{H}$, she can {\em replace} it with any other channel, for example with channel $\bm{G}_1$ by choosing 
\begin{equation}
\bm{Q} = \bm{G}_1 - \bm{H}\,.
\end{equation}
This attack will go undetected, even when Alice and Bob perform the trace check (\ref{tracecon}).

\subsection{Further Attack Strategies}

We indicate here two other possible attack scenarios, whose analysis however is left for future study.

\paragraph{Attack Strategy with Partial \ac{CSI}, Correlated Channels and Conscious Legitimate Terminals} If Alice and Bob knows about the presence of Eve and about the correlation matrices $\bm{K}_1$ and $\bm{K}_2$, they can choose pilot signals that are in the null spaces of the these correlation matrices and then estimate the partial channel in the intersection between the two null spaces. In this case Eve cannot interfere with the channel estimation procedure by any mean. When no null space is available, or the intersection between the two null spaces is empty Alice and Bob can still obtain some advantage by precoding pilots and estimating channels with different weight along different direction. 

\paragraph{Attacks Exploiting Common Exchange of Information} Note that in the proposed solution the estimate of the legitimate channel is not explicitly shared on the common channel. Indeed, the \ac{SK} extraction discloses something about the channel, however this is a very limited information. In principle, Eve could perform multiple attacks by transmitting pilots precoded with various precoding matrices $\bm{Q}$. By analyzing the information shared in public for the key agreement and its variation according to the used precoding matrices she can infer the actual channel $\bm{H}$ between Alice and Bob. Clearly this attack requires time and advanced signal processing capabilities.

\subsection{Comparison with Other Solutions}

As opposed to (the only few) previously proposed methods, the proposed solution relies on pairing based on keys and binary mappings rather than signal processing  approaches \cite{1504.07154v1,6151778,6666096,7136233} that may be prone to errors due to noise and interference. 

When we compare our solution with the technique of \cite{6666096}, some considerations are in place. In \cite{6666096} it is proposed that Bob transmits two consecutive \ac{PSK}-modulated blocks. The first block is a training sequence, while the second comprises random symbols. The attacker knows the training part but not the random part, therefore it either sends other random \ac{PSK} symbols in the second block or it stays silent. In the absence of the attacker, by correlating the two received blocks, a \ac{PSK} symbol must emerge. If the channel is contaminated, no \ac{PSK} symbol appears. Indeed in this case the overhead with respect to our solution is reduced. Still, two observations must be made: a) The phase information may be subject to higher errors than the amplitude information (e.g., due to phase noise), thus fully relying on the phase for security may not be the best option when practical hardware is considered. b) Our solution is based on solid information theoretic principles. 

When compared to \cite{7136233}, our technique presents some advantages. In fact, \cite{7136233} proposes that Bob transmits pilots to Alice, then Alice beamforms to Bob with a gain such that Bob sees an overall channel with predetermined value. The attacker can jam the transmission from Alice to Bob, while listening to the beamforming. As we assume that Eve has as many antennas as Alice, it is then possible for the attacker to know the beamformer used by Alice, thus knowing the channel between Alice and Bob. Once this channel is known, as we have seen, an attack that yields Alice and Bob to estimate $\bm{G}_1$ instead of $\bm{H}$ is possible. Our method overcomes this limitation by not revealing the Alice-Bob channel to the attacker.

\section{Numerical Results}

\begin{figure}[t!]
\includegraphics[width=1\hsize]{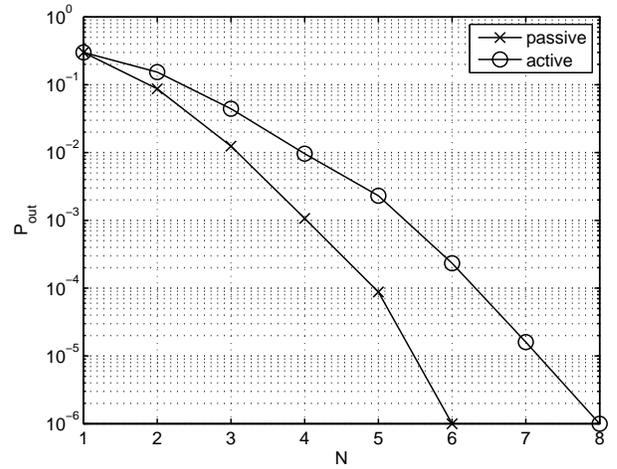}
\caption{\ac{SOP} as a function of the number of antennas $N$, for the case in which Eve does not perform an attack (passive Eve) and the case in which Eve transmits the same pilots of the legitimate transmitters, without precoding (active Eve).}
\label{fig1}
\end{figure}

Focusing on a transmission with rate $R_0$ of which $R_{\rm S} < R_0$, is the secret message rate, the \ac{SOP} can be written as 
\begin{equation}
P_{\rm out}  =  {\rm P}\left[R_0 - \log {\rm det} (\bm{I}_{N} + \bm{G}_1 \bm{B} \bm{B}^H \bm{G}_1^H) < R_{\rm S}\right] 
\label{Pout}
\end{equation}
Note that (\ref{Pout}) only provides the probability that Eve gets some information about the secret message, and not that Bob decodes the message (decodability outage event). For the considered single user scenario we will assume  that Alice uses a beamformer to maximize the rate over Bob's channel, i.e., $\bm{B} = \bm{V}\bm{D}$, where $\bm{U}\bm{\Lambda}\bm{V}^{\rm H} = \bm{H}^{\rm (A)}$ is the \ac{SVD} of the estimated channel and $\bm{D}$ is a diagonal matrix having on the diagonal the square root of the water-filling solution of the channel \acp{SNR} given by $|[\Lambda]_{n,n}|^2$, $n=1, \ldots, N$, under a total unitary power constraint. Note that other beamforming solutions are possible, and their optimization is left for future study.

We assume that channel $\bm{H}$ has \ac{iid} Gaussian standard entries, while channel $\bm{G}$ has zero-mean \ac{iid} Gaussian entries with power 0.5. We have chosen $R_0 = {\rm E}[\log {\rm det} (\bm{I}_{N} + \bm{H} \bm{B} \bm{B}^H \bm{H}^H)]$, under the assumption $\bm{H}^{\rm (A)} = \bm{H}$, and $R_S = 0.2\, R_0$.

Before assessing the performance of the proposed method, we motivate the need of a detection strategy by assessing the \ac{SOP} (\ref{Pout}). Fig. \ref{fig1} shows the \ac{SOP} as a function of the number of antennas $N$ for the two cases where a) Eve is passive, and b) Eve is active and transmits pilots with the same power of Bob. No detection of pilot contamination is performed. We note that an active attacker effectively increases the \ac{SOP}, as expected. Moreover, we note that as the number of antennas increases, the \ac{SOP} tends to zero, as it has been already observed \cite{7248525}.

\begin{figure}
\includegraphics[width=1\hsize]{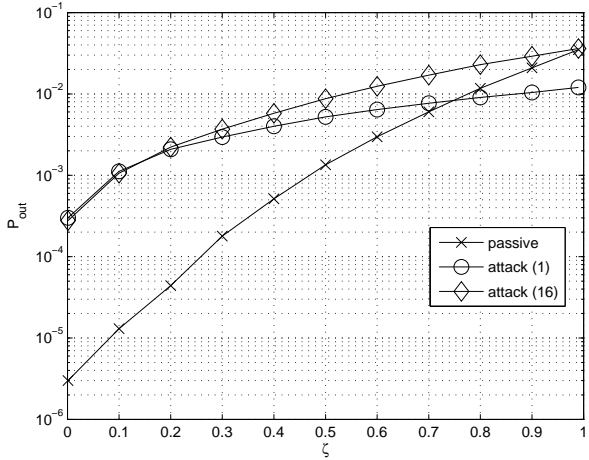}
\caption{\ac{SOP} as a function of the correlation factor $\zeta$, for the case in which Eve does not perform an attack (passive Eve) and the case in which Eve performs the attack reported in Section \ref{corrsec}. $N= 6$ antennas.}
\label{fig2}
\end{figure}

We now focus on the case of Section \ref{corrsec}, where Eve knows channels to Alice and Bob and hence is able to pass the \ac{SK} procedure. However, she has only a statistical knowledge of $\bm{H}$, and channels $\bm{G}_1$ and $\bm{G}_2$ are correlated with $\bm{H}$. Alice is oblivious. As already observed in Section \ref{corrsec}, in this  special scenario, an attack that is able to circumvent both the \ac{SK} procedure and the trace check is possible and the induced channel $\bm{Q}$ that minimizes the \ac{MSE} between the induced channel and $\bm{G}_1$ has been obtained. Therefore, we assess for this case the \ac{SOP}. We assume that $\bm{K}_1 = \bm{K}_2 = \sigma_G^2 \zeta \bm{I}_{N^2}$, $\bm{R}_{12} = \bm{K}_1\bm{K}_2^H$, $\sigma_H^2 = \sigma_G^2 = 0.5$, and $N= 6$ antennas. Fig. \ref{fig2} shows the \ac{SOP} as a function of $\zeta$ for three cases: a) a passive Eve, b) and active Eve using the baseline attack (\ref{baselineattack}), and c) an active Eve using the attack (\ref{expq}). As expected, we observe that as $\zeta$ increases, the \ac{SOP} increases. Moreover, we observe that both active attacks yield a higher \ac{SOP} than when no attack is performed. Lastly, we observe that indeed the choice of $\bm{Q}$ according to (\ref{expq}) yields a further higher \ac{SOP} with respect to the baseline attack.

\section{Conclusions}

We have proposed a technique to contrast the \ac{PCA} in secure \ac{MIMO} systems, based on a key-confirmation procedure. The proposed solution does not reveal the legitimate channel to the adversary, and is effective in rejecting attacks in which the attacker has uncorrelated channel observations with respect to the legitimate channel. An analysis has been performed for a variety of attacks, showing the vulnerabilities when the attacker has access to channel realizations that are correlated to the legitimate channel.

\balance 

\bibliographystyle{IEEEtran}
\bibliography{IEEEabrv,biblio}

\begin{thebibliography}{10}
\providecommand{\url}[1]{#1}
\csname url@samestyle\endcsname
\providecommand{\newblock}{\relax}
\providecommand{\bibinfo}[2]{#2}
\providecommand{\BIBentrySTDinterwordspacing}{\spaceskip=0pt\relax}
\providecommand{\BIBentryALTinterwordstretchfactor}{4}
\providecommand{\BIBentryALTinterwordspacing}{\spaceskip=\fontdimen2\font plus
\BIBentryALTinterwordstretchfactor\fontdimen3\font minus
  \fontdimen4\font\relax}
\providecommand{\BIBforeignlanguage}[2]{{%
\expandafter\ifx\csname l@#1\endcsname\relax
\typeout{** WARNING: IEEEtran.bst: No hyphenation pattern has been}%
\typeout{** loaded for the language `#1'. Using the pattern for}%
\typeout{** the default language instead.}%
\else
\language=\csname l@#1\endcsname
\fi
#2}}
\providecommand{\BIBdecl}{\relax}
\BIBdecl

\bibitem{7270404}
E.~Jorswieck, S.~Tomasin, and A.~Sezgin, ``Broadcasting into the uncertainty:
  Authentication and confidentiality by physical-layer processing,''
  \emph{Proceedings of the IEEE}, vol. 103, no.~10, pp. 1702--1724, Oct 2015.

\bibitem{Shannon}
C.~Shannon, ``Communication theory of secrecy systems,'' \emph{Bell Syst. Tech.
  J.}, vol.~28, pp. 656--715, 1949.

\bibitem{1504.07154v1}
D.~Kapetanovic, G.~Zheng, and F.~Rusek, ``Physical layer security for massive
  {MIMO}: An overview on passive eavesdropping and active attacks,'' \emph{IEEE
  Commun. Mag.}, vol.~53, no.~6, pp. 21--27, June 2015.

\bibitem{6151778}
X.~Zhou, B.~Maham, and A.~Hjorungnes, ``Pilot contamination for active
  eavesdropping,'' \emph{IEEE Trans. Wireless Commun.}, vol.~11, no.~3, pp.
  903--907, March 2012.

\bibitem{6666096}
D.~Kapetanovic, G.~Zheng, K.-K. Wong, and B.~Ottersten, ``Detection of pilot
  contamination attack using random training and massive {MIMO},'' in
  \emph{Proc. IEEE Int. Symp. Personal Indoor and Mobile Radio Commun.
  (PIMRC)}, Sept 2013, pp. 13--18.

\bibitem{7136233}
D.~Kapetanovic, A.~Al-Nahari, A.~Stojanovic, and F.~Rusek, ``Detection of
  active eavesdroppers in massive {MIMO},'' in \emph{Proc. IEEE Int. Symp.
  Personal, Indoor, and Mobile Radio Commun.}, Sept 2014, pp. 585--589.

\bibitem{7248525}
Y.~Wu, R.~Schober, D.~Ng, C.~Xiao, and G.~Caire, ``Secure massive {MIMO}
  transmission in the presence of an active eavesdropper,'' in \emph{Proc. IEEE
  Int. Conf. Commun. (ICC)}, June 2015, pp. 1434--1440.

\bibitem{6675552}
S.~Im, H.~Jeon, J.~Choi, and J.~Ha, ``Robustness of secret key agreement
  protocol with massive {MIMO} under pilot contamination attack,'' in
  \emph{Proc. Int. Conf. on ICT Convergence (ICTC)}, Oct 2013, pp. 1053--1058.

\bibitem{6831343}
------, ``Secret key agreement under an active attack in {MU-TDD} systems with
  large antenna arrays,'' in \emph{Proc. IEEE Global Commun. Conf. (GLOBECOM)},
  Dec 2013, pp. 1849--1855.

\bibitem{Bloch}
M.~Bloch and J.~Barros, \emph{Physical Layer Security}.\hskip 1em plus 0.5em
  minus 0.4em\relax Cambridge University Press, 2011.

\bibitem{7063620}
S.~Tomasin and E.~Jorswieck, ``Pilot-based secret key agreement for reciprocal
  correlated {MIMOME} block fading channels,'' in \emph{Proc. Globecom
  Workshops (GC Wkshps), 2014}, Dec 2014, pp. 1343--1348.

\bibitem{Menezes}
A.~J. Menezes, P.~C. van Oorschot, and S.~A. Vanstone, \emph{Handbook of
  applied cryptography}.\hskip 1em plus 0.5em minus 0.4em\relax CRC Press,
  1997.

\bibitem{6129510}
S.~Nitinawarat and P.~Narayan, ``Secret key generation for correlated
  {Gaussian} sources,'' \emph{IEEE Trans. Inf. Theory}, vol.~58, no.~6, pp.
  3373--3391, June 2012.

\end{thebibliography}

\end{document}